# DBAS: A Deployable Bandwidth Aggregation System

Karim Habak
Department of CS. and Eng.
Egypt-Japan University of
Science and Technology (E-JUST)
Email: karim.habak@ejust.edu.eg

Moustafa Youssef
Department of CS. and Eng.
Egypt-Japan University of
Science and Technology (E-JUST)
Email: moustafa.youssef@ejust.edu.eg

Khaled A. Harras
Computer Science Department
School of Computer Science
Carnegie Mellon University
Email: kharras@cs.cmu.edu

*Abstract*—The explosive increase in data demand coupled with the rapid deployment of various wireless access technologies have led to the increase of number of multi-homed or multi-interface enabled devices. Fully exploiting these interfaces has motivated researchers to propose numerous solutions that aggregate their available bandwidths to increase overall throughput and satisfy the end-user's growing data demand. These solutions, however, have faced a steep deployment barrier that we attempt to overcome in this paper. We propose a Deployable Bandwidth Aggregation System (DBAS) for multi-interface enabled devices. Our system does not introduce any intermediate hardware, modify current operating systems, modify socket implementations, nor require changes to current applications or legacy servers. The DBAS architecture is designed to automatically estimate the characteristics of applications and dynamically schedule various connections or packets to different interfaces. Since our main focus is deployability, we fully implement DBAS on the Windows operating system and evaluate various modes of operation. Our implementation and simulation results show that DBAS achieves throughput gains up to 193% compared to current operating systems, while operating as an out-of-the-box standard Windows executable, highlighting its deployability and ease of use.

## I. INTRODUCTION

The widespread deployment of various wireless technologies coupled with the increase of multi-interface enabled devices are providing users with many alternatives for sending and receiving data. Current devices and operating systems, however, are not exploiting the true potential of these interfaces. Simultaneously leveraging these interfaces by potentially aggregating their bandwidths can lead to higher throughput, improved end-user experience, and efficient resource utilization.

Researchers have addressed the multi-interface bandwidth aggregation problem over the years where solutions and techniques are implemented at different layers of the protocol stack. Application layer solutions typically require the applications to be aware of the existing multiple interfaces and to take the responsibility of utilizing them [1]. Socket level solutions, on the other hand, modify the kernel socket handling functions to enable existing applications to use multi-interfaces [1], [2]. Although [1] does not modify existing applications, it requires changes to the legacy servers in order to support these new sockets. On the other hand, [2] requires feedback from the applications about their performance, and hence is not backwards-compatible with previous versions of the applications. Many bandwidth aggregation techniques, however, naturally lie in the transport layer [3]–[13]. These solutions replace TCP with mechanisms and protocols that handle multi-interfaces. Such techniques require changes to the legacy servers and hence have a huge deployment barrier. Finally, network layer approaches update the network layer to hide the variation in interfaces from the running TCP protocol [14]–[16]. These solutions require proxy servers to connect with or update the network layer protocols in both communicating end-points.

In order to have successful bandwidth aggregation systems for multi-homed or multi-interfaced devices, it is inevitable for these systems to be deployable. We particularly refer to the ability of the system to be installed on a large range of devices with no changes to the existing applications or legacy servers, and without requiring any extra hardware such as proxy servers or special routers. The fact that modern operating systems, such as Windows and Linux, allow users to use only one of the available interfaces, even if multiple of them are connected to the Internet, attests that all the current proposals for bandwidth aggregation face a steep deployment barrier.

We therefore present a Deployable Bandwidth Aggregation System (DBAS) for multi-interface enabled devices. Our system is based on a middleware that lies below the application layer. Unlike previous socket based approaches [1], [2], we do not change the standard sockets but rather intercept standard TCP connection requests and assign the connections to different interfaces in a way transparent to the application. This allows our system to be utilized with no modifications to these applications. DBAS also does not require any update to the operating system kernel implementation and its deployment is as easy running a setup program in the Windows operating system.

The basic mode of DBAS uses a connection-oriented scheduling technique that allows DBAS-enabled clients to quickly communicate with legacy servers without making any changes to them. However, if both communicating end-points are DBAS-enabled, we switch to a packet-oriented scheduling technique that allows us to increase the bandwidth, better handle network conditions, and easily migrate connections in case of failure. Achieving deployability in our approach comes with its own set of challenges including automatically estimating the applications and interfaces characteristics and scheduling the different connections/packets to different interfaces for the two modes of operations. These challenges must also be addressed in a transparent way that operates seamlessly to the user. However, our main contribution is building a deployable system that can be tuned in order to implement more scheduling techniques and target more system metrics.

We evaluate our system via implementation. For further stress testing we conduct further evaluation via NS2 simulations [17]. Currently, our DBAS prototype is deployed using a standard setup programs, highlighting its deployability and ease of use. Our results also quantify the gain of each component in our system as well as the performance differences between connection-oriented and packet-oriented DBAS.

The remainder of this paper is organized as follows. Section II presents the overall architecture of DBAS, its components, and the scheduling techniques we propose. We present the details of the DBAS implementation in Section III and evaluate these scheduling techniques in Section IV. Finally, Section V concludes the paper and provides directions for future work.

## II. A Deployable Bandwidth Aggregation System

In this section, we provide the details of the DBAS system. Figure 1 gives an overview of the system architecture. We consider a client host, which is equipped with multiple network interfaces connected to the Internet. Each of those interfaces has its own parameters in terms of bandwidth, latency, and loss ratio. The device is also running multiple applications with different communication characteristics. DBAS has two modes of operation, depending on whether the other end of the connection supports DBAS or not: a connection-oriented mode and a packet-oriented mode. In the connection-oriented mode, the unit of scheduling is a single connection that can be assigned to one of the available interfaces. Once assigned to an interface, all the packets of this connection utilize the same interface. On the other hand, the packet-oriented mode allows data of the same connection to be routed through different interfaces. This mode, therefore, provides finer scheduling granularity and flexibility. We now describe the various components that contribute to the operation of DBAS.

### A. Mode Detection Module

The purpose of this module is to detect whether the end point of the connection supports DBAS or not to enable the optional packet-oriented mode. This module is implemented as a service that runs on a specific port reserved for DBAS. When a client tries to establish a new connection, it first attempts to open a dummy connection to this port. If successful, this indicates that DBAS is supported at the end point and hence the packet-oriented mode can be used. If so, DBAS will open multiple connections to the destination seamlessly from the application and will distribute the data across those interfaces. Otherwise, DBAS reverts to the connection-oriented mode.

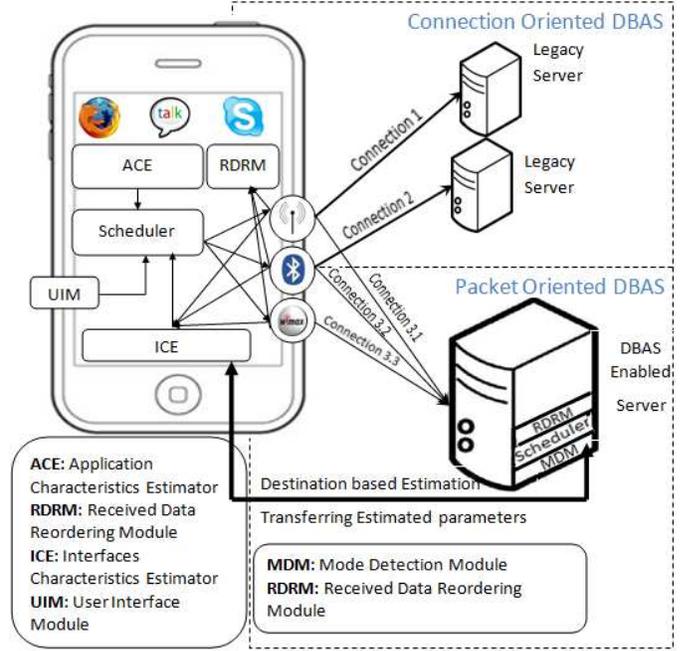

Fig. 1. DBAS system architecture.

To avoid the delay of determining the operation mode, the connection-oriented mode can be used simultaneously while probing the end point.

### B. Application Characteristics Estimator

To be fully deployable, DBAS does not require any changes to existing applications. Knowing the application characteristics, however, enables us to fully utilize the available interfaces and make better scheduling decisions regardless of the mode of operation. Our approach is to automatically estimate the characteristics of the applications based on their behavior. These characteristics are stored in a database for keeping track of the applications behavior. This functionality is the responsibility of the application characteristics estimation module that utilizes both qualitative and quantities measures.

*1) Qualitative Measures:* Some features can be used to characterize the behavior of an applications. For example, the process name can be used to determine whether the process is realtime, e.g. Skype, or bandwidth intensive, e.g. an FTP client. Similarly, specific ports reserved by the application can also be used to characterize the applications behaviors.

*2) Quantitative Measures:* The module also estimates the average connection data demand in bytes of any given application. After a connection is terminated, the module updates the estimated values of connection data demand ($C_{demand}$) as:

$$C_{demand} = (1-\alpha)C_{demand} + \alpha CC_{demand} \quad (1)$$

where $CC_{demand}$ is the number of bytes transmitted by the just terminated connection and $\alpha$ is a smoothing coefficient, taken equal to 0.125 [1], to evaluate the equation efficiently. We note that the granularity of estimation can be rendered more fine grained at the expense of increased complexity and lower scalability.

---

[1]This way, the estimate can be obtained by an efficient shift operation.

### C. Interface Characteristics Estimator

This module is responsible for estimating the characteristics of each network interface. In particular, it estimates the available bandwidth and packet error rate at each interface. This estimation depends on the following two modes of operation.

*1) Connection-Oriented Mode:* In this mode, this module periodically connects and communicates to various geometrically dispersed servers to estimate the uplink and downlink available bandwidth for each interface. These estimates are then combined with the statistics collected during the normal data transfer operation. These estimates are sufficient as the bandwidth bottlenecks typically exist at the client's end not the server's. This assumption is realistic because typically the servers across the Internet are connected to high bandwidth links and designed to scale with many clients.

*2) Packet-Oriented Mode:* In this mode, both ends of the connection implement DBAS. This can be leveraged to obtain more accurate, and route-specific estimates. In this case, the module interacts with the destination to better estimate the interfaces' parameters. Finally, in the packet-oriented mode, a chunk is the unit of scheduling. A chunk represents a group of bytes from the application's data stream. Currently, we set the chunk size to the maximum transmission unit (MTU) of an interface.

Overall, our connection-oriented mode allows DBAS to work with existing applications and legacy server. The packet-oriented mode can then be used to further enhance performance if both end-points are DBAS-enabled. This mode increases throughput, better handles network variations, and is more robust to interface failure.

### D. Scheduler

The scheduler uses the application and interface characteristics estimators to schedule the connections and packets to different interfaces based on the operation mode. We propose and experiment with different connection oriented (CO) and packet oriented (PO) scheduling techniques that provide different complexities and performance tradeoffs as we demonstrate in the evaluation section. The schedulers used are:

*1) Only One Scheduler:* Used to represent schedulers typically used in current operating systems such as Windows or Linux. It assigns connections to only one of the available interfaces.

*2) Optimal Scheduler:* This is a virtual scheduler that can achieve the sum of the throughput of all interfaces. This scheduler represents an upper bound on the performance of any scheduler.

*3) CO Round Robin:* Assigns connections to network interfaces in a rotating basis.

*4) CO Weighted Round Robin:* Assigns connections to network interfaces in a rotating basis weighted by the interface estimated bandwidth such that higher bandwidth interfaces get assigned more connections.

*5) CO Maximum Throughput:* Connections are assigned to the network interface that will maximize the overall system throughput. This occurs by minimizing the the time needed to finish the current system load in addition to the load introduced by the new connection. This scheduler relies on the estimated bandwidth for each interface and the estimated connection data demand per application.

*6) PO Round Robin:* This is the packet-oriented round robin scheduler which assigns packets/chunks to network interfaces in a rotating basis.

*7) PO Weighted Round Robin:* Packets are assigned to network interfaces in a rotating basis weighted by the interfaces estimated bandwidth.

### E. Received Data Reordering Module

This module is utilized in the packet-oriented mode. It is responsible for reordering the incoming data chunks from different interfaces in order to pass them on to the application layer at the receiving end. This is enabled by the sender adding a DBAS header which only contains a chunkId that is used in reordering the chunks when they reach their destination. When an interface is down, unacknowledged chunks can then be resent over other interfaces, showing the effect of connection migration.

### F. User Interface Module

This module is responsible for obtaining the user's preferences and interface usage policies. The user can configure this module to enforce some interface selection criteria. For example, the user may wish to assign specific applications (e.g. realtime applications) to specific interfaces (e.g. wired).

## III. IMPLEMENTATION

To verify the deployability of DBAS, we implement it on the Microsoft Windows Operating System; currently, we have clients for Windows 7, Vista; and XP. We choose the Windows platform to further demonstrate the deployability of DBAS even on closed source operating systems. In this section, we briefly discuss our implementation of DBAS specifically highlighting our DBAS middleware and monitoring application. Both software modules are installed using a standard Windows executable.

### A. DBAS Middleware

We implement the DBAS middleware as a Layered Service Provider (LSP) [18], which is installed as part of the TCP/IP protocol stack in the Windows OS. This middleware implementation uses the same concept used in implementing firewalls and network proxies in order to control traffic flows. The procedure starts when the application, which aims to connect to the Internet, uses the Winsock 2 API. Windows will dynamically link it to the Winsock 2 library which contains the implementation of all the Winsock 2 functions. This library sends its requests to the service provider which later forwards it to the base protocol, such as TCP/IP. Our DBAS LSP intercepts the Winsock 2 API requests and either schedules the connection to its selected interface or distributes data across the different network interfaces based on the mode of operation. In addition, it performs the functionality of the mode detection, application characteristic estimation, interface characteristic estimation, and received data reordering modules described in Section II.

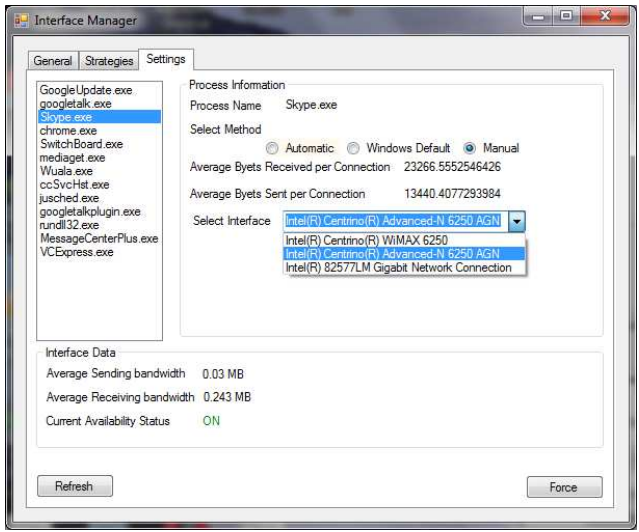

Fig. 2. DBAS monitoring application.

### B. Monitoring Application

The monitoring application, of which a snapshot is shown in Figure 2, represents the user interface module that captures the user's preferences and interface usage policies, and further monitors DBAS behavior. This module is mainly for testing purposes, where it allows the user to select one of the scheduling techniques which DBAS offers. It also provides the ability to perform manual assignment of certain applications to certain interfaces. Finally, this module allows users to monitors DBAS internal data structures estimated values by interfacing with the DBAS middleware.

## IV. PERFORMANCE EVALUATION

In this section we evaluate the performance of DBAS via implementation. Results have been validated using simulation on NS2 [17], but not included for space constraints. We start by describing our experimental setup followed by the results.

### A. Experimental Setup

We use a testbed that consists of three nodes: a server, a client, and an intermediate traffic shaper node. The server is the connection destination that may be DBAS-enabled or not. The intermediate node is a device running the NIST-NET [19] network emulator to emulate the varying network characteristics of each interface. The client is the connection generator enabled with multiple network interfaces. On the client, we run different applications that vary in terms of the number of connections per second they open ($\beta$) and the average connection data demand ($\lambda$). The client is connected to the intermediate node through two different interfaces $IF_1$ and $IF_2$. The server is connected to the intermediate node using a single high bandwidth link $L_1$. We note that the combined bandwidth of $IF_1$ and $IF_2$ is less than the server bandwidth in order to test the true impact of varying the interface characteristics and scheduling strategies.

On the client, we choose to evaluate DBAS using two classes of applications: small load, and large load. The small load applications represent typical web browsing applications and have an average connection data demand of $\lambda_{small} = 22.38KB$ [20]. On the other hand the large load applications represent the P2P and FTP applications that have an average connection data demand of $\lambda_{large} = 0.285MB$ [20]. The connection establishment rate follows a poisson process with mean ($\beta$) connections per second. $\beta_{small}$, and $\beta_{large}$ are changed to achieve different application mixes. Each experiment represents the average of 15 runs.

Throughout our evaluation, we use throughput as our main metric while varying several parameters including application workloads and the various interface characteristics. Table I lists the main parameters we employed in our evaluation.

### B. Results

In this section, we quantify the the gain of each of our components through monitoring its impact on the overall systems performance. We study the impact of interface heterogeneity and application heterogeneity on the performance of the connection-oriented and packet-oriented schedulers we propose. We note that the optimal scheduler performance is the sum of the bandwidths of $IF_1$ and $IF_2$. We do not plot it for the clarity of figures.

*1) Impact of Interface Heterogeneity:* In this experiment, we fix the bandwidth of $IF_1$ at 2Mbps and loss rate to 0% and vary $IF_2$ settings as shown in Table I. The main goal of this experiment is to quantify the gain of the interface characteristics estimator. The application mix is the default shown in Table I.

**Bandwidth:** Figure 3 shows the impact of changing the bandwidth of the second interface on the overall system throughput. For the connection-oriented schedulers (Figure 3(a)), when the bandwidth of $IF_2$ is very low compared to $IF_1$, using only $IF_1$ outperforms the round robin and weighted round robin schedulers. The main reason of this is that they do not take the connection loads into account in their scheduling decisions. On the other hand, since the maximum throughput scheduler considers the connection load and characteristics in its decision, it outperforms the other connection-oriented

TABLE I
EXPERIMENTS PARAMETERS.

| Parameter | Value range | Nominal Value(s) |
|---|---|---|
| $L_1$ (Server) bandwidth (Mbps) | 6 | 6 |
| $L_1$ (Server) packet loss ratio (%) | 0 | 0 |
| $IF_1$ bandwidth (Mbps) | 2 | 2 |
| $IF_1$ packet loss ratio (%) | 0 | 0 |
| $IF_2$ bandwidth (Mbps) | 0.25 - 2 | 1, 2 |
| $IF_2$ packet loss ratio (%) | 0 - 10 | 0 |
| $\beta_{small}$ (connections/sec) | 13 | 13 |
| $\beta_{large}$ (connections/sec) | 0 - 5 | 1 |

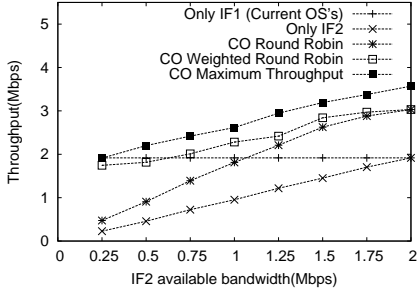
(a) Connection-oriented schedulers

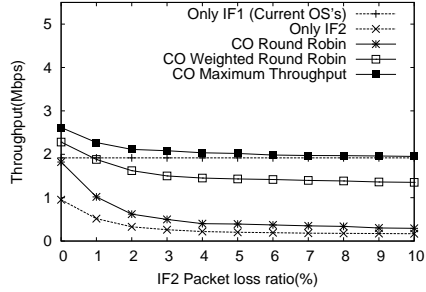
(a) Connection-oriented schedulers

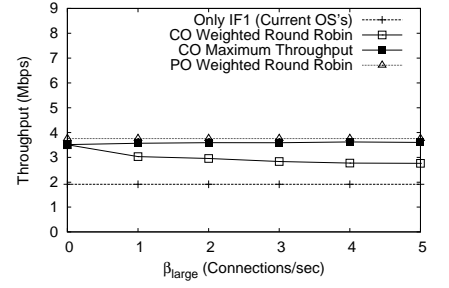
Fig. 5 Effect of applications heterogeneity.

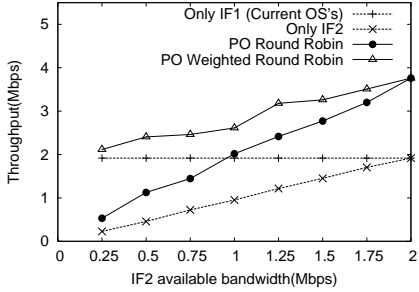
(b) Packet-oriented schedulers

Fig. 3. Impact of interfaces heterogeneity (bandwidth).

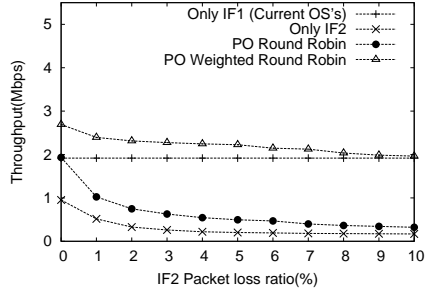
(b) Packet-oriented schedulers

Fig. 4. Impact of interfaces heterogeneity (packet loss ratio).

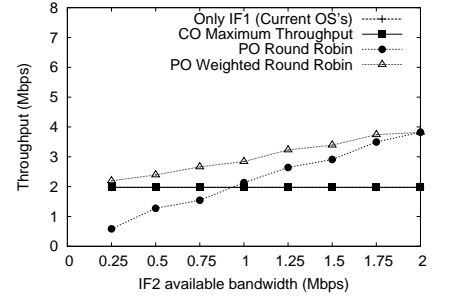
Fig. 6. Effect of scheduling granularity.

approaches. The figure also shows that weighted round robin achieved throughput gains up to 369% compared to round robin since it takes advantage of the interface characteristics estimator module.

Similarly, for the packet-oriented schedulers (Figure 3(b)), PO weighted round robin achieves the highest throughput. The fine-grained scheduling of packets makes the effect of the system under-utilization less severe compared to the CO schedulers. As we discuss later, PO schedulers are not sensitive to the applications characteristics.

**Loss ratio:** Figure 4 shows the impact of changing the loss ratio of $IF_2$ on the throughput using different schedulers. $IF_2$ bandwidth was fixed at 1Mbps. The relative behavior of the schedulers is basically similar to that of changing the bandwidth of the interface. The non-linear behavior of the throughput curve is due to TCP AIMD policy; once TCP detects a loss, it drops its congestion window exponentially which severely impacts the sending rate.

*2) Impact of Application Heterogeneity:* Figure 5 shows the impact of the system workload on performance. For this experiment, we fix $\beta_{small}$ at 13 connections/sec while varying $\beta_{large}$. The figure shows that packet-oriented scheduling techniques are not sensitive to the application connection characteristics since they operate on the packet level. On the other hand, the CO weighted round robin achieves lower throughput than the CO maximum throughput scheduler because it does not make use of the application characteristics estimator module; the interface estimator module essentially leads to throughput gains of up to 130%.

*3) Scheduling Granularity:* We now demonstrate the advantage of using fine grained packet-oriented scheduling when available by comparing the best CO scheduler, i.e. the CO maximum throughput scheduler, to the best PO scheduler, i.e. the PO weighted round robin scheduler. In these experiments, we introduce a single long-lived connection running from the beginning coupled with several short-lived connections ($\beta_{large} = 0.25$). Again, the two network interfaces were set to have a 0% packet loss ratio and a 2Mbps bandwidth.

Figure 6 shows that, in this extreme case, the throughput achieved by the CO schedulers is the same as using only a single interface. This is due to their coarse scheduling granularity which cannot distribute the packets of the single long-lived connection across multi-interfaces. On the other hand, the fine-grained scheduling granularity of the packet-oriented schedulers better adapts with such scenarios. PO weighted round robin achieves the highest throughput because it distributes the packets of the single long lived connection over all interfaces. PO round robin performs poorly since it does not take interface heterogeneity into account. The throughput gains highlight the use of the mode detection module as well as the received data reordering module.

### C. Discussion

Beyond the tradeoffs observed in our results, there are a few points we believe are worth highlighting.

First, our results show that leveraging multi-interfaces on mobile devices significantly enhances the overall throughput. Packet-oriented schedulers, due to their fine-grained scheduling, achieve a near-optimal throughput (9% difference in the worst case). Connection oriented schedulers can generally achieve this near-optimal performance. However, in extreme cases, such as the one described in Section IV-B3, they deviate from the optimal with their lower bound on throughout still as good as the current operating systems performance. This is due to their coarse-grained scheduling as a connection is assigned to one interface from its creation to termination. They show also the more they make use of our system components the better performance they gain.

Second, our results also demonstrate the importance of interface and application characteristics estimation in multi-interface schedulers. The CO round robin and the PO round robin schedulers do not take the interface characteristics into account in their decisions, which renders their performance worse than the single interface case (Figure 3) when the system is under-utilized. Similarly, it is important to take the application characteristics into account in the CO schedulers. If not, as in the CO weighted round robin scheduler, a large connection can be assigned to a slow interface degrading the system performance making it even worse than the single interface case (Figures 5,6).

The CO maximum throughput scheduler takes interface and application characteristics into account. Therefore, its performance is the best among all CO schedulers. The PO weighted round robin schedulers takes the interface characteristics into account, making it the best PO scheduler, having a near-optimal performance under all cases. We note, however, that packet-oriented schedulers are insensitive to application characteristics: their packet-level scheduling makes all connections arriving from the applications appear as a pool of packets (chunks) to DBAS. In such cases, the application characteristic estimation module is not required.

We finally observe that the DBAS architecture makes each DBAS-enabled client independent from each other. This enables DBAS to be highly scalable in terms of the number of clients. The impact of increasing the number of DBAS-enabled nodes in the same network is sharing the common bandwidth, which will be captured by the interface characteristics estimation module. The extra required capacity has to be provisioned for by the service provider. We have confirmed these observations via simulations.

## V. Conclusion and Future Work

In this paper we presented DBAS, a deployable bandwidth aggregation system for devices with multiple network interfaces. We have presented the components of our system, demonstrated its deployability, compared it to other state-of-the-art solutions, and evaluated its performance via implementation. We have introduced two scheduling classes in DBAS that perform connection level and packet level scheduling across networking interfaces; tradeoffs in their performance have been clearly demonstrated. Overall, DBAS exploits the available bandwidth of all interfaces and is backwards compatible with legacy servers and clients.

We are currently expanding our system in different directions including having DBAS schedulers adapt to other system metrics such as cost and power. Adaptive scheduling strategies sensitive to dynamic user profiles and needs represent milestones are also worth pursuing. Finally, we plan to make our code public for the research community and implement our system on the Linux and Android.